\documentclass{rQUF2e}

\usepackage{epstopdf}
\usepackage{subfigure}
\usepackage{bbm}
\usepackage{amsmath}
\usepackage{algorithm}
\usepackage[noend]{algpseudocode}
\usepackage{caption}
\usepackage{bm}

\makeatletter
\def\BState{\State\hskip-\ALG@thistlm}
\makeatother

\theoremstyle{plain}

\theoremstyle{definition}

\theoremstyle{remark}

\usepackage{xcolor}

\usepackage{color,soul}

\begin{document}


\title{Forecasting market states}

\author{ PIER FRANCESCO PROCACCI $^1$ and TOMASO ASTE $^{1,2}$\\
\affil{$^1$Department of Computer Science, UCL, Gower Street, London, WC1E6BT, UK\\
$^2$Systemic Risk Centre, London School of Economics and Political Sciences, London,  WC2A 2AE, UK\
} \received{v1.1 released June 2018} }

\maketitle

\begin{abstract}
We propose a novel methodology to define, analyse and forecast market states. 
In our approach  market states are identified by a reference sparse precision matrix and a vector of  expectation values.
In our procedure each multivariate observation is associated to a given market state accordingly to a minimisation of a penalized Mahalanobis distance. 
The procedure is made computationally very efficient and can be used with a large number of assets. 
We demonstrate that this procedure is successfull at clustering different states of the markets in an unsupervised manner.  In particular, we describe an experiment with one hundred log-returns and two states in which the methodology automatically associates states prevalently to pre- and post- crisis periods with one state gathering periods with average positive returns and the other state  periods with average negative returns, therefore discovering spontaneously the common classification of `bull' and `bear' markets.
In another experiment, with again one hundred log-returns and two states, we demonstrate that this procedure can be efficiently used to forecast off-sample future market states with significant prediction accuracy. This methodology opens  the way to a range of applications in risk management and trading strategies in the context where the correlation structure plays a central role.
\end{abstract}


\begin{classcode}C38, G61, G15, G17 \end{classcode}

\section{Introduction}
Markets do not always behave in the same way. 
In common terminology, there are periods of `bull' market in which prices are more likely to rise and periods of `bear' market in which prices are more likely to fall. These different `states' of markets are commonly attributed in literature to unobservable, or latent, regimes representing a set of macroeconomic, market and sentiment variables. \\

Many time series models presented in literature tried to capture this phenomenon. Among the most popular methods, it is worth mentioning the TAR models \citep{Tong}, trying to estimate `structural breaks' in the time series process, and the Markov Switching models \citep{Hamilton89}, where the change in regimes are parametrized by means of an unobserved state variable typically modelled as Markov chain. However, the application of TAR models in finance is frequently criticized since it cannot be established with certainty when a structural break has occurred in economic time series and the prior knowledge of major economic events could lead to bias in inference \citep{campbell_97}. Markov switching models, on the other hand, are highly affected by the curse of dimensionality. In particular, for slightly more complex dynamics than the original proposal \citep{Hamilton89}, we need to rely on variational inference techniques or MCMC methods \citep{Tsay05, Kim}. This implies that, in a multivariate context and particularly if we aim to extract information on the switching from the correlation structure, estimation becomes difficult to perform.

Other approaches focus on clustering of observations into groups: `similar' data objects are discovered on the basis of some criteria for comparisons. Most works related to clustering of time series are classified into two categories: subsequence time series clustering and point clustering. Subsequence clustering involves the clustering of sliding windows of data points and usually aim at discover repeated patterns. Example are Dynamic Time Warping \citep{Liao}, Hierarchical methods \citep{Nevill} or pattern discovery \citep{Ren}.
In point clustering methods, instead, each multivariate observation at each time instance $t$ is assigned to a cluster. In most popular approaches, however, this is done based on a distance metric \citep{Grabarnik2001, Fabozzi2004, Zolhavarieh14, Hendricks2016, Hallac2016}.\\

In a multivariate context,  different `states' of markets are not only reflected in the gains and losses, but also in the relative dynamics of prices. Indeed, the correlation structure changes between bull and bear periods indicating that there are structural differences in these market states. 
Most common approaches in the industry assume -for convenience- a stationary correlation structure \citep{Duffie7, Black92}. 
However, it is well established that correlations among stocks are not constant over time \citep{Lin94, Ang99, Aste_Correlation} and increase substantially in periods of high market volatility, with, asymmetrically, larger increases for downward moves (see, for example, \citep{Ang02, Cizeau2010, Schmitt13}).
Indeed, various approaches have been proposed in literature to model and predict time-varying correlations. Examples are, for instance, the generalized autoregressive conditional heteroskedasticity (GARCH) models  by \citep{bollerslev1990modelling} or  the Dynamic Conditional Correlation (DCC) model by \citep{Engle12}. However, most of these models are not able to cope with more than a few assets due to the curse of dimensionality having numbers of parameters that increases super-linearly with the number of variables \citep{Danielsson11}.
Other approaches have been focusing on the study of changes in a time-varying correlation matrix computed from a rolling window. 
This is, for instance, the case of estimators like the RiskMetrics \citep{Riskmetrics96} or \citep{Lee01}. 
However, since these approaches use only a small part of the data, these estimators have large variances and, in case of high dimensionality, may lead to inconclusive estimates \citep{Laloux99}. \\
\cite{Hallac17} introduced a clustering algorithm called TICC (Toeplitz Inverse Covariance Clustering), originally proposed for electric vehicles, where classification into states is constructed from a likelihood measure associate with a referential sparse precision matrix (inverse covariance matrix). Instead of considering each observation in isolation, however, in their approach they cluster short subsequences of observations so that the covariance matrix constructed on the subsequences provides a representation of the cross-time partial correlations. In this setting, then, by imposing a Toeplitz constraint to the precision matrix of each regime, the cross-time partial correlations are constrained to be constant and, hence, covariance-stationarity is enforced. 
This method has a number of appealing features from a financial perspective, although the structure of data considered by the authors is significantly different from noisy data in finance.
\\ \\
In this paper we build on \citep{Hallac17} and propose a similar Covariance based Clustering. However, we consider single observations and do not enforce Toeplitz structure on the precision matrix. We, therefore, call this methodology ICC - Inverse Covariance Clustering.
Analogously to \citep{Hallac17}, we also enforce temporal coherence by penalizing frequent switches between market states and favouring temporal consistency. 
Another difference is that we do not directly maximise likelihood, but rather we assign states to clusters accordingly to their Mahalanobis distance \citep{de2000mahalanobis}. 
We experiment with this methodology in the context of financial time series and provide a detailed analysis of the role played by sparsity and temporal consistency, while assessing the significance of the clusters. Finally, we show that the cluster classification can be used for one step ahead off-sample prediction.

Our approach simplify and clarify the definition of `market state' by identifying each state with a sparse precision matrix and a vector of expectation values which are associated to a set of multivariate observations clustered together accordingly with a given procedure.
In the following, the precision matrix of market state `$k$' is denoted with $\bm J_k$ and it represents the structure of partial correlations between the system's variables. In the multivariate normal case, two nodes are conditionally independent if and only if the corresponding element of $\bm J_k$ is equal to zero.
A sparse  precision matrix provides an easily interpretable and intuitive structure of the market state with all the most relevant dependencies directly interconnected in a sparse network.
Furthermore, sparsity reduces the number of parameters from order $n^2$ (with $n$ the number of variables) to order $n$ preventing overfitting \citep{lauritzen1996} and filtering out noisy correlations \citep{Aste_Parsimonious,Musmeci16_multiplex}. \\

The {\bf segmentation procedure} 
starts by setting the number of clusters $K$ (in the present paper we limit to $K=2$) and assigns multivariate observations to  clusters randomly. From these $K$ sets of data we compute the sample means $\bm{\mu}_k$ and the precision matrices $ \bm{J}_k$ and we then  iteratively re-assign points to the cluster with smallest 
\begin{equation}
  \mathcal{M}_{t,k} =  d^2_{t,k} +   \gamma \mathbbm{1} \lbrace \mathcal K_{t-1}\not= k \rbrace \;\;.
  \label{likelihood_def0}
\end{equation}
where  $\bm{X}_t=[x_{t,1},x_{t,2},...,x_{t,n}]$  is the $n$-stocks multivariate observation at time $t\;(=1,...,T)$; $\bm{\mu}_k$ is the vector of the means for cluster $k$; $\bm{J}_k$ is the (sparse) precision matrix for cluster $k$; $  d^2_{t,k} = (\bm{X}_t-\bm{\mu}_k)^T \bm{J}_k \;(\bm{X}_t-\bm{\mu}_k)$ is the  the square Mahalanobis distance of observation $\bm{X}_t$ in cluster $k$ with respect to the cluster centroid $\bm{\mu}_k$;  $\gamma$ is a parameter penalizing state switching;  $\mathcal K_{t-1}$ is the cluster assignment of the observation at time $t-1$.
We considered as well clustering with respect to maximum likelihood and minimum Euclidean distance, however we report only about the procedure with Mahalanobis distance which is the one providing the best results. 
Specifically,  Euclidean distance is very efficient in distinguishing positive and negative returns but does not distinguish well between pre- and post-crysisis periods. 
The maximum likelihood instead identifies very well the crisis period but then it is much less clean in classifying the `bull' and `bear' market states.
Let us note that the used Mahalanobis distance clustering is producing high likelihood although not maximal.

{The clustering assignment procedure is made computationally efficient by using the Viterbi algorithm \citep{Viterbi67, Bishop} that transforms an otherwise $O(K^T)$  procedure into $O(KT)$ (Appendix \ref{Viterbi_app}).}
Further, the sparse precision matrix $\bm{J}_k$ is computed efficiently from the observations in each cluster by means of  the TMFG-LoGo network filtering approach \citep{AsteTMFG,Aste_Parsimonious}. TMFG-LoGo approach  has proven to be more efficient and better performing, particularly when few data are available \citep{Aste_Parsimonious,aste2017causality}, with respect to other techniques  such as GLASSO \citep{Friedman_GLasso}.
Implementation has been performed with and in-house built-for-purpose python package. 
This is the first time this methodology is introduced and applied to financial data  and market states analytics. 

In this paper we report results for two experiments performed over a {\bf dataset} of daily closing prices of $n = 2490$ US stocks entering among the constituents of the Russel 1000 index ($RIY \; index$) traded between 02/01/1995 and 31/12/2015. For each asset $i = 1, ..., n$, we calculated the corresponding daily log-returns  $r_{i}(t) = \log(P_{i}(t)) - \log(P_{i}(t-1))$, where $P_{i}(t)$ is the closing price of stock $i$ at time $t$.

\section{Clustering}
As mentioned in the introduction, our primary goal is to efficiently cluster noisy, multivariate time series into meaningful regimes, while controlling for temporal consistency.
In this first experiment, we considered the entire dataset between  02/01/1995 and 31/12/2015 and estimated two referential market states. In order to explore the role of each building block of our algorithm and to compare it to a traditional baseline method, we investigate five models:
\begin{itemize}
\item a) ICC Model - Sparse precision matrix and temporal consistency
\item b) ICC Model - Full precision matrix and temporal consistency
\item c) ICC Model - Sparse precision matrix
\item d) ICC Model - Full precision matrix
\item e) Gaussian Mixture Model - Full Covariance
\end{itemize}
Model (a) is the present proposed ICC methodology. Model (b) considers full precision matrices $\bm{J}_k$ instead of sparse ones. Model (c) relaxes temporal consistency allowing for $\gamma = 0$ in Eq. \ref{likelihood_def0}. Model (d) has $\gamma = 0$ full precision matrices. Finally, Model (e) is a conventional Gaussian Mixture Model \citep{Bishop} that has been chosen as a baseline method given the similarities with the ICC approach. We analysed and compared the resulting clusters both in terms of market properties to which the two clusters are associated and in terms of temporal consistency.
First, we focused on a subset of 100 stocks chosen at random among those that have been continuously traded throughout the observed period. Random choice of the basket is to avoid selection bias. We then consider random resamplings to assess the robustness when different stocks are considered.
\begin{figure}[h]
\begin{center}
\begin{minipage}{140mm}
\begin{center}
\subfigure[Time series segmentation results]{
\resizebox*{6cm}{!}{\includegraphics{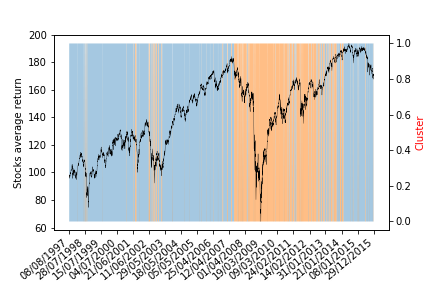}}\label{sample-figure_part_a}}
\subfigure[Mean (\textit{left}) and standard deviation (\textit{right}) of each stock for each temporal cluster]{
\resizebox*{6cm}{!}{\includegraphics{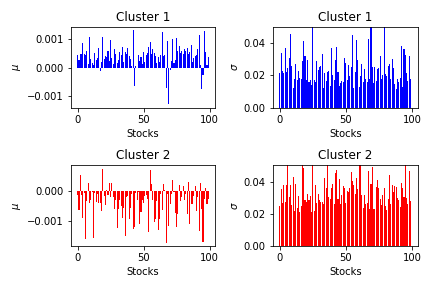}}\label{sample-figure_part_b}}
\caption{Clustering segmentation for experiment 1 over the whole dataset.  Panel (a) reports the cumulative average return at each time $t$ across the 100 stocks;  in this picture, the blue background corresponds to time instances assigned to Cluster 1 and the orange background correspond instead to time instances assigned to Cluster 2.
Panel (b) reports mean and standard deviation of each of the 100 stocks respectively computed using the returns assigned to each of the 2 clusters. We observe that Cluster 1 exhibits positive mean returns (\textit{`bull'} state) and lower levels of volatility for all the considered stocks, while for cluster 2  all the stocks present negative mean returns (\textit{`bear'} state) and higher levels of volatility.
\label{clustering}}
\end{center}
\end{minipage}
\end{center}
\end{figure}

We optimized the temporal consistency parameter by grid-searching as described in Appendix \ref{Viterbi_app} and  {used $\gamma = 16$ for ICC Sparse (a) and $\gamma =14.7$ for ICC Full (b)} in both the experiments presented in this paper.
The two referential precision matrices, $\bm J_1$ and $\bm J_2$,  obtained with this experiment had  { 344 non-zero entries (dependency network edges) of which 142 } were common to both states showing a good level of differentiation, but also significant overlaps between the two market states.
The number of points assigned to each cluster were respectively  { 3295 for cluster 1 and 1704 } for cluster 2.
Figure \ref{clustering} reports with colored background the points' assignment for the two clusters. We can observe there is a good spatial consistency. For instance, the average number of consecutive days in cluster  { 1 is 25.3 days}. We also note that cluster 1 (blue background) tend to be associated with periods of rising market prices whereas cluster 2 (orange background) appears more present during crisis and market downturns.
We indeed discovered that -automatically- the methodology assigns \textit{`bull'} market periods (positive mean returns) to  cluster 1 and \textit{`bear'} market periods (negative mean returns) to cluster 2. We can for instance observe in Figure \ref{sample-figure_part_a}  that  { 52 consecutive observations during the 2001-2002 $.com$ bubble crisis and 211 consecutive observations} during the 2007-2008 global financial crisis have been assigned to the \textit{bear} cluster 2.  From Fig.\ref{sample-figure_part_b} we observe that the \textit{bull} cluster 1 has, indeed, average positive returns for all stocks whereas the \textit{bear} cluster 2 has average negative returns. Furthermore, also the standard deviations are different between the two cluster assignments. 

\begin{figure}
\begin{center}
\begin{minipage}{140mm}
\begin{center}
\includegraphics[scale=0.50]{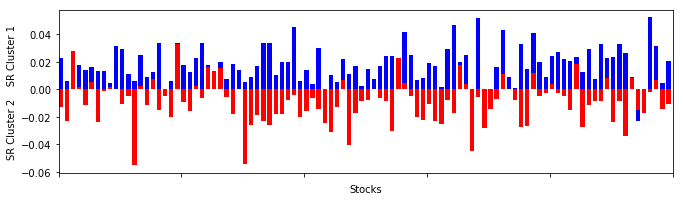}
\caption{Estimated Sharpe Ratio (SR) for each of the 100 stocks in the sample. The blue bars report the SR computed from log-returns in Cluster 1, whereas the red bars report the SR computed from log-returns in Cluster 2.
\label{Sharpe_fig}}
\end{center}
\end{minipage}
\end{center}
\end{figure}

To compare the two clusters on a risk-adjusted basis, we computed the Sharpe ratio \citep{Sharpe66,Sharpe94} for each stock in each cluster. We found for the \textit{bull} cluster an average annualized Sharpe ratio equal to  { $1.2$, with $5^{th}$ and $95^{th}$ percentiles respectively equal to $0.84$ and $1.78$,} while the \textit{bear} cluster had average  {$-0.96$, with $-1.03$ and $-0.24$} as $5^{th}$ and $95^{th}$ percentiles. It is therefore clear that the two clusters have very different risk-return profiles. Figure \ref{Sharpe_fig} reports the  Sharpe ratios in the two clusters for the 100 stocks. 
In order to verify robustness and generality of the results we computed the same quantities for 100 other randomly chosen baskets of 100 stocks. 
For all resampled baskets of stocks we found a consistent clusterization in \textit{bull} and \textit{bear} regimes with Sharpe ratios for at least 75\% of stocks  larger than zero for the bull  state and significantly smaller than zero for the bear  state.   
Across the 100 resamplings, the two clusters had average number of elements respectively equal to  { 3451 and 1293.} \\ 

\par 
\textbf{Sparsity and Temporal Consistency.}
In order to assess the role of sparsity and temporal consistency, we performed the same analysis on the `alternative' ICC Models (b)-(d) and the GMM (e).

Table \ref{model_comp} summarizes  {the number of stocks having positive/negative Sharpe ratio in both clusters over 100 resamplings. In the table, each couple refers to the number of  stocks having positive SR in \textit{bull} (left) and negative SR in \textit{bear} (right) states. We found that, in absence of temporal consistency constraints, both the ICC models (c, d) meaningfully classify clusters with and without sparsity. However, when temporal consistency is considered, ICC Full (b) is significantly affected by the constraint while ICC Sparse (a) provides robust results. GMM delivered the worst clusters in terms of risk/return significance}.

\begin{center}
\begin{minipage}{140mm}
\begin{center}
\resizebox*{11cm}{!}{\begin{tabular}{lccc}
&\textbf{Median}&\textbf{$5^{th}$ percentile}&\textbf{$95^{th}$ percentile}\\ \cline{2-4}
\textbf{$GMM$}&(69,64)&(48,53)&(75,81)\\
\textbf{$ICC$ Full, $\gamma = 0$}&(77,78)&(67,71)&(92,98)\\
\textbf{$ICC$  Sparse, $\gamma = 0$}&(85,87)&(69,75)&(96,95)\\
\textbf{$ICC$  Full, $\gamma = 14.7$}&(73,74)&(68,65)&(78,80)\\
\textbf{$ICC$  Sparse, $\gamma = 16$}&(75,81)&(65,69)&(86,90)\\
\end{tabular}}
\captionof{table}{Positive/Negative Sharpe ratio for (`\textit{bull'},`\textit{bear'}) states. Median, $5^{th}$ and $95^{th}$ percentiles obtained from 100 random resamples of the stocks composing the dataset.\label{model_comp}}
\end{center}
\end{minipage}
\end{center}

Focusing on temporal consistency, Table \ref{temporal} reports the number of switches and the segment length resulting from the cluster assignments of the five models. When no temporal consistency is enforced (c,d), ICC provides the less temporal consistent results with small differences related to sparsity. This also explains the good results obtained by the models in terms of risk/return significance.
When constrained to be temporal consistent, ICC Full (b) shows large variability in temporal consistency across samples with some having only a few switches over the whole period and others having several hundreds. ICC Sparse (a) is instead more consistent with a few hundred switches over the whole period which  {are less than 1/3 of the switches in GMM (e).}

\begin{figure}[H]
\begin{center}
\begin{minipage}{140mm}
\begin{center}
\subfigure{
\resizebox*{11cm}{!}{\begin{tabular}{lccc}
& &\textbf{Number of Switches}& \\ \cline{2-4}
&\textbf{Median}&\textbf{$5^{th}$ percentile}&\textbf{$95^{th}$ percentile}\\ \cline{2-4}
\textbf{$GMM$}&785&540&874\\
\textbf{$ICC$ Full, $\gamma = 0$}&1203&992&2176\\
\textbf{$ICC$  Sparse, $\gamma = 0$}&1157&727&1421\\
\textbf{$ICC$  Full, $\gamma = 14.7$}&204&54&306\\
\textbf{$ICC$  Sparse, $\gamma = 16$}&208&120&298\\
\end{tabular}}}
\subfigure{
\resizebox*{11cm}{!}{\begin{tabular}{lccc}
& &\textbf{Segment length}& \\ \cline{2-4}
&\textbf{Median}&\textbf{$5^{th}$ percentile}&\textbf{$95^{th}$ percentile}\\ \cline{2-4}
\textbf{$GMM$}&5.07&2.4&11.8\\
\textbf{$ICC$ Full, $\gamma = 0$}&3.3&1.68&4.38\\
\textbf{$ICC$  Sparse, $\gamma = 0$}&3.5&2.8&6.65\\
\textbf{$ICC$  Full, $\gamma = 14.7$}&22.64&14.6&38.26\\
\textbf{$ICC$  Sparse, $\gamma = 16$}&23.6&18&55.27\\
\end{tabular}}}
\captionof{table}{Temporal consistency metrics. Number of switchings and Segment lengths over 100 resampligs.\label{temporal}}
\end{center}
\end{minipage}
\end{center}
\end{figure}

\section{Role of sparsity}
\begin{figure}[H]
\begin{center}
\begin{minipage}{140mm}
\centering
\resizebox*{11cm}{!}{  \includegraphics[width=1\linewidth]{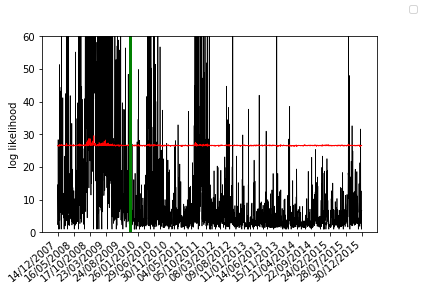}}
\caption{Train and test log likelihood observation-wise using TMFG (red line) and Ridge (black line) precision matrices. The green vertical line divides train and test set. Ridge peaks reach values outside the range up to 320.}
\label{TMFG_tests}
\end{minipage}
\end{center}
\end{figure}

In previous works \citep{Aste_Parsimonious}, the TMFG-LoGo approach has proven to perform better than other filtering approaches including GLasso and Ridge providing the additional advantages of efficiency and fixed sparsity level with no need to calibrate hyperparameters \citep{AsteTMFG}. In this Section we motivate the choice of TMFG-LoGo filtering procedure in terms of statistical significance by comparing the performances of the TMFG-LoGo to the cross-validated Ridge \textit{l}$_2$ penalized inverse covariance (Ridge) on our dataset. We considered the widely used Ridge penalization as robust estimate of the empirical inverse covariance matrix and compared it to TMFG-LoGo and show that, when applied to our dataset, TMFG-LoGo produces more stable likelihood results than Ridge.
We used  { 40\% of the data (from 31/12/2007 to 31/12/2015)} as test set, and we considered as train sets the $q$ observations preceding the  { test set (until 30/12/2007).} 
The penalization parameter of Ridge was defined by cross validating within the train set. 
To compare  TMFG-LoGo and the cross-validated Ridge we computed the log-likelihoods $\mathcal{L}_{s,k} = 1/2 ( \log| \mathbf J_k | - d^2_{s,k} - p \log(2\pi) )$ using the two covariance estimates and compared them.
Figure \ref{TMFG_tests} shows the likelihood observation-wise computed in train and in test using the TMFG-LoGo and Ridge precision matrices estimated over $q=500$ observations. The TMFG-LoGo likelihoods are much more stable over time suggesting that the procedure was successful in filtering out noise. Table \ref{TMFG_Ridge_table} reports details on mean, $5^{th}$ and $95^{th}$ percentiles of the likelihoods computed in the train and test set. As previously mentioned, TMFG-LoGo likelihoods are much more stable with $5^{th}$ and $95^{th}$ varying a few percent only for TMFG-LoGo and instead varying of more than one order of magnitude in Ridge. We found similar results for TMFG-LoGo and Ridge when different values of $q$ are considered. Note that Ridge log likelihoods have large differences between train and test. This is a typical indication of overfitting. Conversely, TMFG presents small differences indicating that the LoGo procedure acts as a topological penalize.

\begin{figure}[H]
\begin{center}
\begin{minipage}{140mm}
\begin{center}
\subfigure{
\resizebox*{10cm}{!}{\begin{tabular}{lccc}
& &\textbf{Train set}& \\ \cline{2-4}
&Average	&	$5^{th} percentile$	&	$95^{th} percentile$\\\cline{2-4}
$\mathcal{L}_{Ridge}$	&	41.70	&	2.19	&	188.85\\
$\mathcal{L}_{TMFG}$	&	26.71	&	26.53	&	27.22\\
\end{tabular}}}
\subfigure{\resizebox*{10cm}{!}{\begin{tabular}{lccc}
& &\textbf{Test Set}& \\ \cline{2-4}
&Average	&	$5^{th} percentile$	&	$95^{th} percentile$\\\cline{2-4}
$\mathcal{L}_{Ridge}$	&	8.08	&	1.39	&	27.64\\
$\mathcal{L}_{TMFG}$	&	26.55	&	26.44	&	26.73\\
\end{tabular}}}
\captionof{table}{TMFG and Ridge log likelihood metrics - means, $5^{th}$ and $95^{th}$ percentiles - computed in train (top panel) and test (bottom panel) set. TMFG and Ridge precision matrices are estimated using $q=500$ observations.\label{TMFG_Ridge_table}}
\end{center}
\end{minipage}
\end{center}
\end{figure}

\section{Forecasting} \label{For_Sec}
In the second experiment we used our methodology to forecast future states of the market form previous observations.
\begin{figure}[h]
\begin{center}
\begin{minipage}{140mm}
\begin{center}
\includegraphics[scale=0.55]{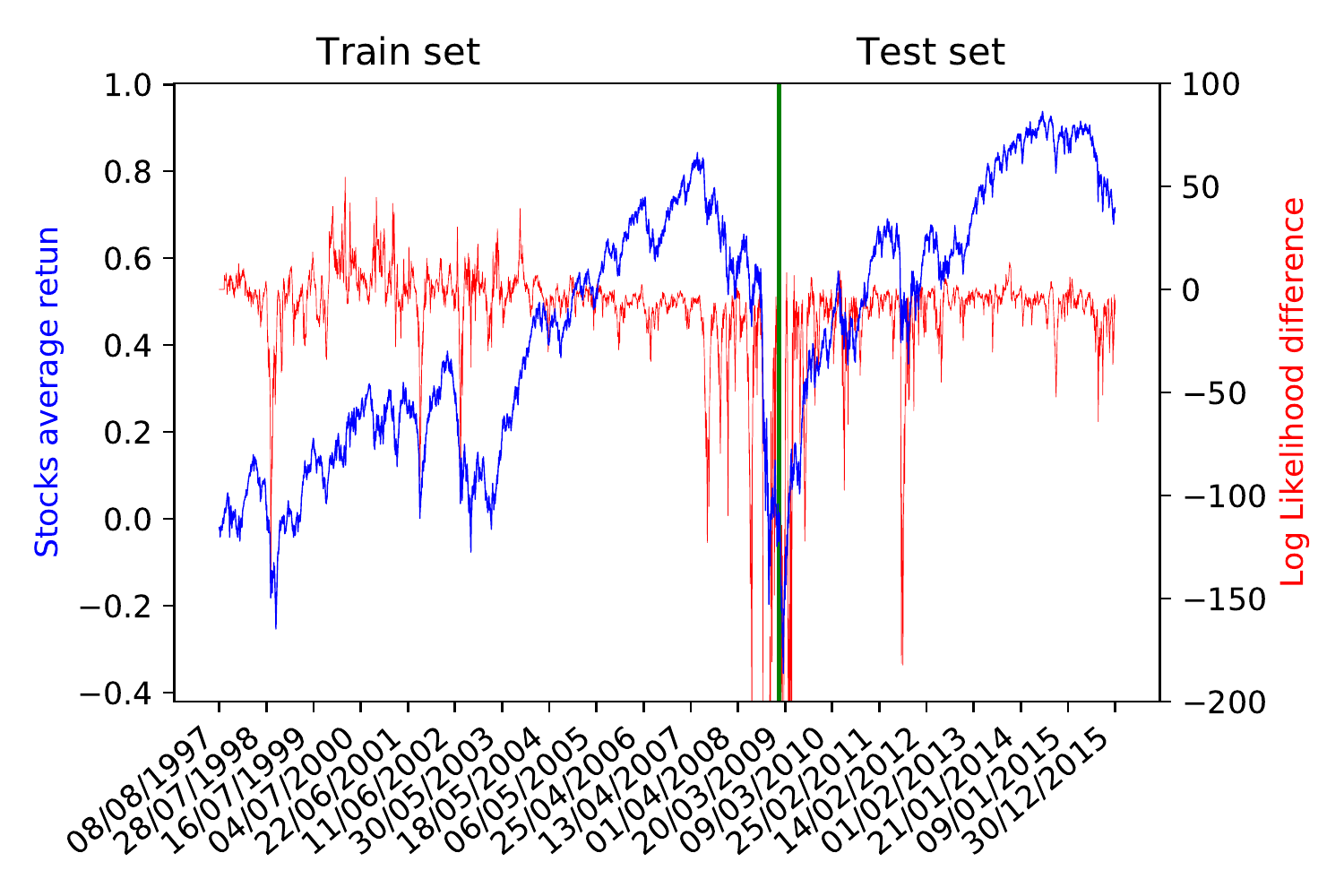}
\caption{Log likelihood ratio and mean returns across train and test sets. The log likelihood ratio of the two states $\mathcal{R}_t$ was computed using using the $\Delta=28\;days$. The green vertical bar indicates the end of the train set and the beginning of the test set. We estimated $\bm{J_1}$ and $\bm{J_2}$ in train and held it fixed for the computation of $\mathcal{R}_t$ also in the test set. The black horizontal line identifies  $\mathcal{R}_t=0$ level, \textit{i.e.} the level above which the \textit{bull} state is more likely. Coherently with previous findings, we can identify persistent market states with a more frequent bull market and regions of bear market.
\label{Likelihood_ratio}}
\end{center}
\end{minipage}
\end{center}
\end{figure}
To this end, we used the first  {65\% of the data (from 02/01/1995 to 05/02/2009) as train set} from which we extracted the two referential precision matrices and means ($\bm J_1$, $\bm{\mu}_1$) and ($\bm J_2$, $\bm{\mu}_2$) (note they are different form the ones of the first experiment in which we used the entire dataset instead). We then forecasted the probability that, given an observation at time $t$, the observation at a following time $t+h$ would belong to state $k$.
This is achieved by performing a logistic regression using the log likelihood ratio of the two clusters  \citep{Neyman33} from a rolling window of length $\Delta$: 
\begin{equation}
  \mathcal{R}_{t} = \sum_{s=t-\Delta+1}^{t} \mathcal{L}_{s,1} - \mathcal{L}_{s,2} \;\;,
  \label{likelihood_rolling_}
\end{equation}
where $\mathcal{L}_{s,k} = 1/2 ( \log| \mathbf J_k | - d^2_{s,k} - p \log(2\pi) )$ is the log-likelihood of observation $\bm{X}_s$ when associated with cluster $k=$ 1 or 2.
In our experiment,  {we considered $\Delta=24$ days, since this is the average length of segments obtained from ICC (a) in the first experiment.} Figure \ref{Likelihood_ratio} provides a visual representation of the likelihood ratio computed for each cluster and of its evolution as compared to market movements. The green vertical line divides the train set form the test set.  The logistic regression of market states $\mathcal K_{t}$ against the log likelihood ratio $\mathcal{R}_t$ can be written as
\begin{equation}
  P(\mathcal K_{t+h} =1,2 \,|\; \mathcal{R}_t=x) = \frac{1}{1+e^{-\left(\beta_0 + \beta_1 x\right)}} \;\;,
  \label{Logistic_eq}
\end{equation}
where the parameters ${\beta_0}$ and ${\beta_1}$ are estimated through maximum likelihood \citep{Bishop}. 
We estimated all parameters ($\bm{J}_1$, $\bm{J}_2$, $\bm{\mu}_1$, $\bm{\mu}_2$, $\gamma$, ${\beta_0}$ and ${\beta_1}$) in the train set and estimated a threshold or cut-off point of 0.54 by cross-validation in the train set. We then used these parameters to predict, in the test set, the next day state given the log-likelihood ratio $\mathcal{R}_{t}=x$. 
Specifically, we predict  { $\widehat{\mathcal K}_{t+1}=1$ if $P(\mathcal K_{t+1} =1 \,|\; \mathcal{R}_{t}=x)>0.54$ and  $\widehat{\mathcal K}_{t+1}=2$ otherwise.} 
For instance,  { for the day 30-Mar-2010 (test set) we predicted a \textit{bull} state with probability $P(\mathcal K_{30-Sep} =1 \,|\; \mathcal{R}_{29-Mar})=0.77$, where $\mathcal{R}_{29-Mar}$ was computed using the observations from 06-Mar to 29-Mar-2010 ($\Delta=24$ days, all in the test set) and the parameters $\bm{\mu}_k$, $\bm{J}_k$, $\gamma$, ${\beta_0}$ and ${\beta_1}$ were the ones calibrated on the train set with data until 31/04/2009.}

To assess the goodness of our approach we compared test set predictions with the classification performed over the whole period in the first experiment (see Fig.\ref{clustering}).
We used three metrics \citep{Hastie14} to assess the performance of our classification method: the True Positive Rate $TPR$ (number of elements correctly assigned to cluster 1 divided by total number of elements in cluster 1),  the True Negative Rate $TNR$ (number of elements correctly assigned to cluster 2 divided by total number of elements in cluster 2) and  Accuracy $ACC$ (number of correct predictions in cluster 1 or 2 divided by  total number of elements).
In order to test for the robustness of our method, we randomly resampled the 100 stocks  and performed the classification experiment considering the new dataset. 
We repeated this process 100 times and stored the three performance metrics $TPR$, $TNR$ and $ACC$. Table \ref{Performance} presents a summary of the results obtained. A good level of $ACC$ is obtained across resamplings with only the 5th percentile falling slightly below $50\%$. As we can see, $TPR$ is higher than $50\%$ at the $5^{th}$  percentile while $TNR$ presents a good median result, but a low $5^{th}$  percentile showing that it can be difficult to correctly forecast. This indicates that there is a tendency to over-assign time-instances to cluster 1 (\textit{bull} state) and conversely missing predictions for the less frequent \textit{bear} state. 
Nonetheless, we verified (by using the hypergeometric distribution as reported in \citep{aste2017causality}) that these $TNR$ are statistically significant at 0.01 level indicating that there is, indeed, significant prediction power also for the bear state. 
Let us stress that the present forecasting exercise is not optimized and there are several ways these performances can be improved. However, this is beyond the purpose of the present paper where we privileged simplicity over performances.
\begin{center}
\begin{minipage}{140mm}
\begin{center}
\begin{tabular}{lccc} 
&\textbf{Median}&\textbf{$5^{th}$ percentile}&\textbf{$95^{th}$ percentile}\\ \cline{2-4}
\textbf{$TPR$}&0.68&0.51&0.93\\
\textbf{$TNR$}&0.52&0.39&0.78\\
\textbf{$ACC$}&0.54&0.47&0.69\\ 
\label{Performance}
\end{tabular}
\captionof{table}{Out-of-sample performance metrics using the ICC log likelihood ratio as independent variable. Median, $5^{th}$ and $95^{th}$ percentiles obtained from 100 random resamples of the stocks composing the dataset.} 
\end{center}
\end{minipage}
\end{center}

To compare the previous results with a baseline method, we estimated the logistic regression in Eq. \ref{Logistic_eq} using the fraction of stocks that at time $t-1$ were presenting positive returns as independent variable. Aim being to compare our ICC log-likelihood to a much simplified version of the the information about the correlation structure.
Same estimation scheme is used and a treshold of 0.61 is obtained by cross validation. Results are reported in Table \ref{Performance_fraction}.  {While this simplified information still provides a median accuracy close to 50\%, the model has overall inferior performances with respect to the ICC log likelihood ratio case reported in} Table \ref{Performance}.

\begin{center}
\begin{minipage}{140mm}
\begin{center}
\begin{tabular}{lccc}
&\textbf{Median}&\textbf{$5^{th}$ percentile}&\textbf{$95^{th}$ percentile}\\ \cline{2-4}
\textbf{$TPR$}&0.71&0.67&1\\
\textbf{$TNR$}&0.24&0.0&0.77\\
\textbf{$ACC$}&0.47&0.38&0.62\\
\end{tabular}
\captionof{table}{Out-of-sample performance metrics using the fraction of positive strocks as independent variable. Median, $5^{th}$ and $95^{th}$ percentiles obtained from 100 random resamples of the stocks composing the dataset. \label{Performance_fraction}}
\end{center}
\end{minipage}
\end{center}

\section{Conclusions}
In this paper  we presented a novel methodology to define, identify, classify and forecast market states.
In addition to accuracy, intuitiveness and forecasting power, our procedure is numerically very efficient and able to process high dimensional datasets.
We reported two experiments to illustrate that the method is efficient and reliable in identifying and predicting accurate and interpretable structures in multivariate, non-stationary financial datasets. 
These two examples use two clusters and 100 variables, however we verified that analogous results hold for larger or smaller numbers of variables and similarly interesting classifications emerge also when  three or more clusters are used.
The choice of two clusters has been only motivated by simplicity. The fact that they turned out to be respectively populated mostly with average positive and negative returns associated with pre- and post-crisis periods was unexpected by us and opens potentials for completely novel ways to use multivariate analytics for the forecasting of stock market returns. This also greatly simplified the interpretation of these states as \textit{`bull'} and \textit{`bear'} markets. Of course, in reality, there are more than two market states  and common definition of bull and bear markets are often blurry.
In this work we did not attempt to optimize results favouring, instead, simplicity and interpretability and, therefore, there is a large open domain of exploration to refine the methodology.
We also adopted several methodological choices that can be modified in future works. For instance, the segmentation with the Mahalanobis distance turned out to be a powerful tool in the reported experiments, however there is a broad range of possible metrics for clustering and experiments with Euclidean distance or Likelihood also produce interesting results. Further, the choice of TMFG network over other possible information filtering networks or other sparsification methodologies can be investigated. Temporal consistency could had also being performed differently by using a hidden Markov model approach (see note in Appendix \ref{Viterbi_app}). The choice of logistic regression to forecast market state is just one simple possibility among many regression options that might make better use of the information content of our regimes' structures.
All these and other methodological choices have been motivated by simplicity and intuitiveness.
Since one of the main achievement of our methodology is computational efficiency allowing to apply the methodology to high dimensional datasets, further work will include new sources of information (e.g. news, economic indicators, sentiment).

\bibliographystyle{rQUF}

\newpage
\raggedbottom
\break

\appendices

\section{The Viterbi algorithm} \label{Viterbi_app}
\begin{center}
\begin{figure}[h]
\begin{center}
\begin{minipage}{140mm}
\begin{center}
\includegraphics[scale=0.5]{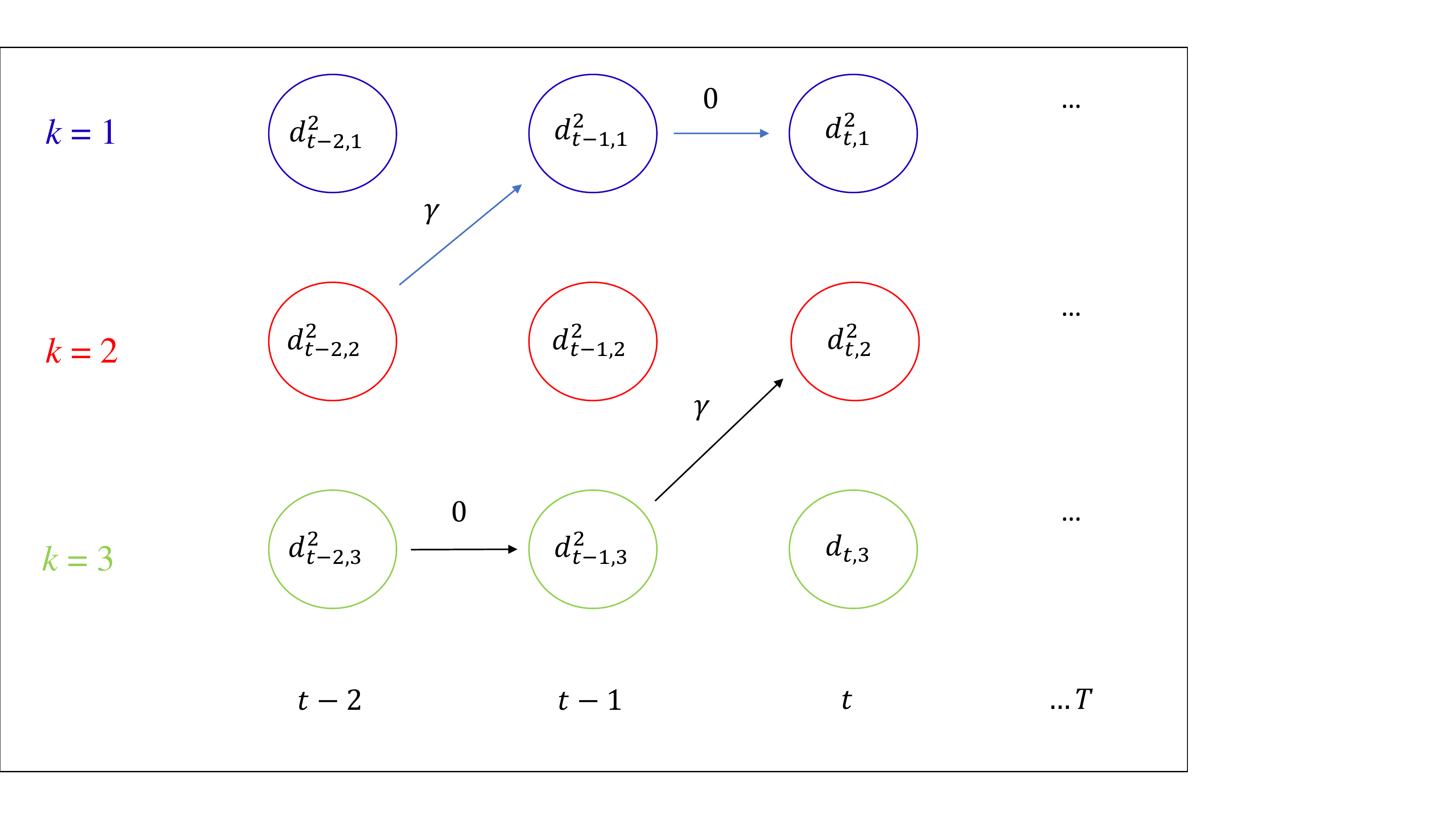}
\caption{Example of two among the $K^T$ possible paths considering $K=3$ clusters and $T$ observations. $\mathcal{L}_{t,j}$ represents the log likelihood of the multivariate observation at time $t$ if assigned to cluster $j$. If an observation is assigned to same cluster as the previous one, no penalty is applied, otherwise a \textit{cost} weighted by the parameter $\gamma$ is added.
\label{Viterbi_prob}}
\end{center}
\end{minipage}
\end{center}
\end{figure}
\end{center}

Figure \ref{Viterbi_prob} provides a visualization of the problem of assigning points to clusters. 
Based on the parameters estimates ($\bm{\mu}_k$ and $\bm{J}_k$ via TMFG-LoGo), we compute the Mahalanobis distance of every multivariate observation obtaining, for each cluster $k$ and for each observation $t$, a value  $d^2_{t,k} = (\bm{X}_t-\bm{\mu}_k)^T \bm{J}_k \;(\bm{X}_t-\bm{\mu}_k) $.

We need to consider the best \textit{sequence} of latent states which is not the set of best individual states. In particular, if we introduce a cost parameter $\gamma$ that penalizes cluster switching, the problem complexity becomes combinatorial, since we need to account for the whole sequence or \textit{path} of assignations. In particular, given $K$ potential cluster assignment of $T$ points (multivariate observations), the number of potential paths grows exponentially with the length of the chain to $K^T$ possible assignments of points to clusters. Based on a dynamic programming approach, the Viterbi algorithm \citep{Viterbi67} provides an efficient solution with complexity $O(KT)$ (\textit{i.e., linear}) to this problem, searching the space of the paths and finding the most efficient path. The Viterbi algorithm in the convenient formulation by \citep{Hallac17} is sketched in \ref{ViterbiAlgo}.

\begin{algorithm}[h]
\fontsize{10pt}{10pt}\selectfont
\begin{minipage}{140mm}
\begin{center}
\caption{Viterbi algorithm}\label{ViterbiAlgo}
\begin{algorithmic} 
\State \textbf{Input}
\State $d^2_{t,k}=\,$ square Mahalanobis distance of observation $t$ if assigned to state $k$
\State $\gamma=\,$ time consistency parameter
\State
\State \textbf{Initialize}
\State\hspace{\algorithmicindent} PreviousCost = array of K zeros
\State\hspace{\algorithmicindent} CurrentCost = array of K zeros
\State\hspace{\algorithmicindent} PreviousPath = array of K elements
\State\hspace{\algorithmicindent} CurrentPath = array of K elements

\State
\For {each observation $t = 1,...,T$} 
\For {each state $k = 1,...,K$} 
\State MinVal = index of minimum value of PreviousCost
\If {PreviousCost$\left[MinVal\right]+\gamma>$ PreviousCost$\left[k\right]$}
\State CurrentCost$\left[k\right]=\,$PreviousCost$\left[k\right] + d^2_{t,k}$
\State CurrentPath$\left[k\right]=\,$PreviousPath$\left[k\right]$.append$\left[k\right]$
\Else
\State CurrentCost$\left[k\right]=\,$PreviousCost$\left[MinVal\right] +\gamma + d^2_{t,k}$
\State CurrentPath$\left[k\right]=\,$PreviousPath$\left[MinVal\right]$.append$\left[k\right]$
\EndIf
\State PreviousCost=CurrentCost
\State PreviousPath=CurrentPath
\EndFor
\EndFor
\State FinalMinVal=index of minimum value of CurrCost
\State FinalPath=CurrPath[FinalMinVal]
\end{algorithmic}
\end{center}
\end{minipage}
\end{algorithm}

A more general formulation can be implemented by describing the paths as Markov chains and introducing a transition probability between the states. 
However, under the Markov chain formalism the expression in Eq.\ref{likelihood_rolling_} for the likelihood ratio is no longer consistent because it implies iid multivariate observations.

\raggedbottom
\break

\end{document}